# Achieving High Capacity Transmission With $N$-Dimensional Quasi-Fractal UCA

Hongyun Jin, *Student Member, IEEE*, Wenchi Cheng, *Senior Member, IEEE*,
Haiyue Jing, *Student Member, IEEE*, Jingqing Wang *Member, IEEE*, and Wei Zhang, *Fellow, IEEE*

*Abstract*—The vortex electromagnetic wave carried by multiple orthogonal orbital angular momentum (OAM) modes in the same frequency band can be applied to the field of wireless communications, which greatly increases the spectrum efficiency. The uniform circular array (UCA) is widely used to generate and receive vortex electromagnetic waves with multiple OAM-modes. However, the maximum number of orthogonal OAM-modes based on UCA is usually limited to the number of array-elements of the UCA antenna, leaving how to utilize more OAM-modes to achieve higher channel capacity with a fixed number of array-elements as an intriguing question. In this paper, we propose an $N$-dimensional quasi-fractal UCA ($N$D QF-UCA) antenna structure in different fractal geometry layouts to break through the limits of array-elements number on OAM-modes number. We develop the $N$-dimensional OAM modulation ($N$OM) and demodulation ($N$OD) schemes for OAM multiplexing transmission with the OAM-modes number exceeding the array-elements number, which is beyond the traditional concept of multiple antenna based wireless communications. Then, we investigate different dimensional multiplexing transmission schemes based on the corresponding QF-UCA antenna structure with various array-element layouts and evaluate the optimal layout type and dimension to obtain the highest channel capacity with a fixed number of array-elements. Simulation results show that our proposed schemes can obtain a higher spectrum efficiency, surpassing those of alternative array-element layouts of QF-UCA and the traditional multiple antenna systems.

## I. Introduction

FOR the sixth-generation (6G) wireless system application scenarios, there is a demand for enhanced channel capacity and higher spectrum efficiency (SE) [1]–[3]. Electromagnetic waves propagate carrying both linear momentum and angular momentum, where angular momentum includes spin angular momentum and orbital angular momentum (OAM). Theoretically, vortex electromagnetic waves carrying OAM can potentially contain an infinite number of mutually orthogonal eigenstates, referred to as OAM-modes in this paper [4]–[6]. Vortex electromagnetic wave technology, as a new mode division multiplexing technology [7] [8], has received significant researach attention in the field of wireless communications [9]–[12]. Utilizing vortex electromagnetic wave transmission techniques with mutually orthogonal OAM-modes, the performance of communication systems can be greatly enhanced without increasing the current spectrum bandwidth, thus leading to a significant increase in the SE [13]–[16].

The uniform circular array (UCA), as the classical structure to generate and receive vortex waves with multiple OAM-modes, offers notable advantages in terms of flexibility and convenience in generating and receiving multiple OAM-modes [17]. The uniformly arranged circular array-elements are fed with a continuous step phase, resulting in the acquisition of vortex waves with a continuous phase shift along the array-elements [18]–[20]. The authors of [21] generated OAM beams with UCA fed by in-phase and verified the generation of different OAM-modes on an experimental platform. The authors of [22] proposed a radial UCA for OAM generation and dual-mode communication based on a multilayer design and gave a theoretical derivation of radial UCA for OAM generation. The authors of [23] investigated the OAM-mode of vortex wave generated by UCA, derived theoretical equations for the radiation field, and determined the factors affecting the pattern distribution. Then, an experimental setup of a vortex radio beam was created and the radiation field was measured to verify the theoretical results.

The design of the vortex electromagnetic wave multiplexing transmission holds significant importance in enhancing the channel capacity [24]–[29]. The authors of [30] proposed the OAM embedded multiple-input-multiple-output (MIMO) communication system with UCA antenna to obtain the SE gain for joint OAM and massive-MIMO based wireless communications. The authors of [31] proposed a general scheme for multi-carrier and multi-mode OAM communication based on UCAs, which reduces the burden of estimating channel matrices to better realize the transmission and reception of vortex electromagnetic waves. The authors of [32] aimed to convert singular UCA into concentric UCAs, where multiple parallel low-order OAM-modes can be used to achieve higher channel capacity. The authors of [33] focused on regulating the divergence angles of all non-orthogonal OAM beams to be the same, thus circumventing the issue that large beam divergence of high-order orthogonal OAM-modes results in low channel capacity. The authors of [34] proposed a small-scale circular phased array antenna for OAM-carrying radio beams. Meanwhile, the generation of multiple pure or mixed OAM beams with helical phase fronts is also presented, where the superposition of multiple OAM-modes provides more

H. Jin, W. Cheng, H. Jing, and J. Wang are with the State Key Laboratory of Integrated Services Networks, Xidian University, Xi'an, 710071, China (e-mails: hongyunjin@stu.xidian.edu.cn, wccheng@xidian.edu.cn, hyjing@stu.xidian.edu.cn, and jqwangxd@xidian.edu.cn).

W. Zhang is with School of Electrical Engineering and Telecommunications, the University of New South Wales, Sydney, Australia (e-mail: w.zhang@unsw.edu.au).

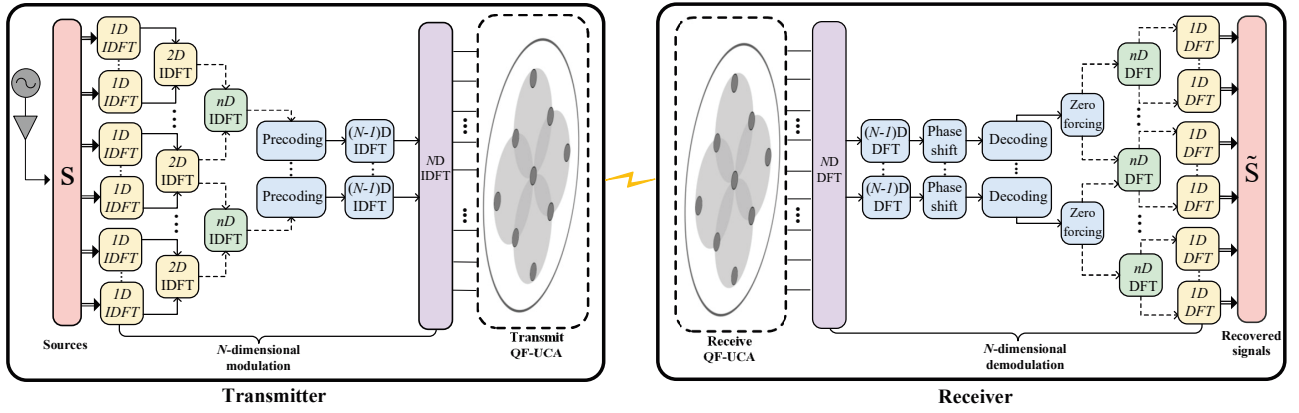

Fig. 1. The system model for $N$-dimensional QF-UCA based OAM multiplexing transmission.

possibilities to enhance the channel capacity.

However, in order to increase the number of available OAM-modes, conventional vortex electromagnetic waves focus on adopting high-order OAM-modes, which leads to a few problems [5]. On the one hand, high-order OAM-modes require more transmitter antenna array-elements and larger transmitter antenna aperture. On the other hand, due to the hollow properties of the vortex electromagnetic wave, the vortex electromagnetic wave with high-order OAM-modes disperses seriously [35], thereby limiting the feasibility of acquiring an adequate number of vortex electromagnetic wave beams for demodulation out of the information at the receiving antenna, which is not conducive to the enhancement of the channel capacity [36]–[38]. How to effectively utilize the aperture of the transmitter antenna to design the layouts of array-elements for transmitting vortex electromagnetic wave represents a significant research challenge. This endeavor aims to highlight the great advantage of the vortex electromagnetic wave over the traditional plane wave, receiving considerable research attention.

In order to solve the above problems, we design a new antenna geometry layout and provide an $N$-dimensional ($N$D) OAM multiplexing method to generate more orthogonal OAM-modes, where the number of OAM-modes is greater than that of array-elements [2]. First, we design an $N$D quasi-fractal UCA (QF-UCA) antenna layout in which each dimensional QF-UCA is taken into account as an integral unit. There are shared array-elements between two adjacent UCAs in QF-UCA, which can generate or receive different OAM-modes in the two adjacent UCAs, respectively. Vortex electromagnetic wave transmission based on conventional single-loop antennas intends to increase the number of available OAM-modes but are limited by the number of array-elements and antenna aperture. The utilization of shared array-elements is able to break through the limitations of the number of array-elements on the number of OAM-modes, and provide a new solution to increase the number of available OAM-modes when the antenna aperture and the number of array-elements are fixed, which is beyond the traditional idea of multiple antennas based wireless communications [39]. Then, we develop $N$-dimensional OAM modulation ($N$OM) and demodulation ($N$OD) schemes based on discrete Fourier transformation (DFT) for multiple OAM-modes transmission. We utilize $N$ separate sets of OAM-modes within $N$D QF-UCA to achieve $N$D OAM multiplexing. Vortex electromagnetic waves for high-capacity transmission can adopt the multiple dimensional modulation and demodulation scheme to further multiplex OAM-modes, which is a novel multiplexing transmission scheme. In addition, we investigate different types of array-element layouts for $N$D QF-UCA and evaluate the optimal layout and dimension to obtain the maximum number of OAM-modes with a fixed number of array-elements. To assess the SEs of OAM multiplexing transmission based on QF-UCA with different numbers of array-elements, we define the average SE of each array-element when the power is uniformly distributed across each OAM-mode as the efficiency of array-element layouts (EOAL) in QF-UCA. Simulation results demonstrate that the optimal array-element layout of QF-UCA can be obtained using our proposed schemes to achieve maximum SE for OAM multiplexing transmission.

The rest of this paper is organized as follows. In Section II, the system model and geometric model of the QF-UCA antenna are provided. In Section III, the $N$OM and $N$OD schemes and the equivalent channel model are analyzed. Section IV provides the numerical results. Finally, we conclude this paper in Section V.

Notation: Matrices and vectors are denoted by the capital letters and the lowercase letters in bold, respectively. The notation blkdiag $(\boldsymbol{X})$ represents a diagonal block matrix with $\boldsymbol{X}$ as its main diagonal elements. The notations $(\cdot)^H$ and $(\cdot)^T$ denote the Hermitian and the transpose of a matrix or a vector, respectively.

## II. THE SYSTEM MODEL FOR $N$D QF-UCA BASED OAM MULTIPLEXING TRANSMISSION

Figure 1 depicts the system model for $N$D QF-UCA based OAM multiplexing transmission. The transmitter includes $n$D ($n \in [1, N]$) inverse discrete Fourier transformation (IDFT) modulation, precoding, and transmit QF-UCA antenna. The 1D, $\cdots$, and $N$D IDFT forms the $N$D OAM modulation for input signals. The transmit QF-UCA antenna is used for emitting multiple OAM-modes signals. The receiver includes receive QF-UCA antenna, $n$D DFT ($n\in[1,N]$) demodulation, phase shift, decoding, and zero forcing. The receive QF-UCA antenna is used for receiving OAM-modes signals. The 1D, $\cdots$, and $N$D DFT forms the $N$D OAM demodulation for received signals. The precoding and decoding are designed to ensure that the OAM-modes are orthogonal at the receiver. The phase shift converts the antisymmetric elements of the equivalent channel matrix into symmetric elements by easily inverting the signal phase, then decoding and zero forcing are performed. The signals carrying different OAM-modes are recovered after $n$D ($n\in[1,N]$) OAM demodulation based on the orthogonality among the OAM-modes. We denote by one-dimensional (1D) UCA the traditional single-loop UCA. For OAM modulation, taking the 1D UCA based OAM modulation as an example, the phase gradient $2\pi l_1/K_1$ is given to the signals in the same branch, where $l_1$ ($0 \leq l_1 \leq K_1 - 1$) denotes the order of OAM-mode used for the 1D modulation and the constant factor $2\pi/K_1$ matches the azimuth difference of any two adjacent array-elements within the 1D UCA. This process can be regarded as applying the 1D IDFT to the input signals. The $n$D OAM modulation is performed that the signals corresponding to the $n$D QF-UCA cells are with a phase gradient $2\pi l_n/K_n$, where $l_n$ ($0 \leq l_n \leq K_n - 1$) denotes the order of OAM-mode used for the $n$D QF-UCA modulation and the constant $2\pi/K_n$ matches the azimuth difference of any two adjacent $(n-1)$D QF-UCA cells. This process can be regarded as applying the $n$D IDFT to the signals. The receiver applies the corresponding $n$D ($n\in[1,N]$) DFT for OAM demodulation.

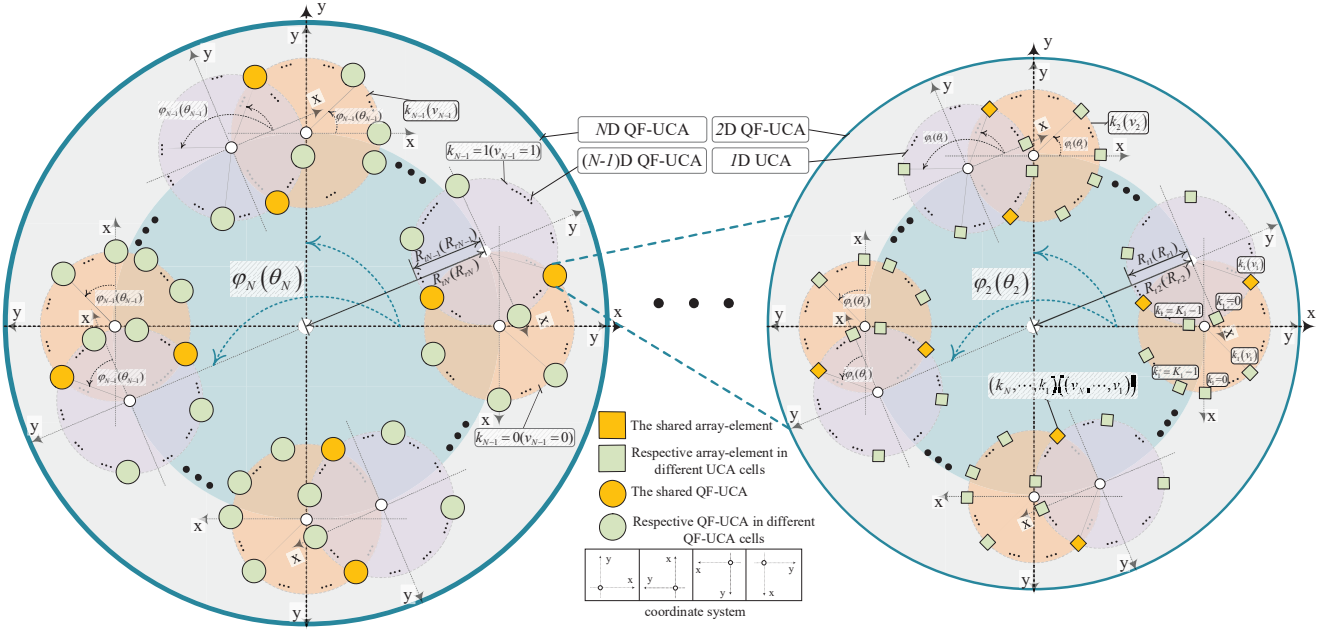

Fig. 2. The $N$D QF-UCA.

Figure 2 depicts the $N$D QF-UCA. In this paper, we assume that the transmit and receive $n$D ($n\in[1,N]$) QF-UCA antennas are strictly aligned with each other and that the design-related requirements for antennas, such as the impedance matching and the spatial correlation, are well satisfied. The transmit $N$D QF-UCA antenna has $N_t$ array-elements, which are divided into $K_N$ $(N-1)$D QF-UCA cells. Then, each $(N-1)$D QF-UCA cells are divided into $K_{N-1}$ $(N-2)$D QF-UCA cells, $\cdots$, each $n$D QF-UCA cells are divided into $K_n$ $(n-1)$D QF-UCA cells, $\cdots$, each 2D QF-UCA cells are divided into $K_2$ 1D UCA cells and $K_1$ array-elements are equipped in each 1D UCA cell. Thus, the $N$D QF-UCA antenna can transmit $\prod_{n=1}^{N} K_n$ data streams. Since there exist shared array-elements among different UCA cells, the total number of transmitted data streams is larger than the number of array-elements equipped in the $N$D QF-UCA, i.e., $\prod_{n=1}^{N} K_n > N_t$. The receive $N$D QF-UCA antenna has $N_r$ array-elements, which $n$D QF-UCA ($n \in (1,N]$) cells are divided into $V_n$ $(n-1)$D QF-UCA cells and $V_1$ array-elements are equipped in each 1D UCA cell. Therefore, the $N$D QF-UCA antenna can receive $\prod_{n=1}^{N} V_n$ ($\prod_{n=1}^{N} V_n > N_r$) data streams. The radius of the $n$D ($n\in[1,N]$) QF-UCA is the distance from the center of the $n$D QF-UCA to the center of the $(n-1)$D QF-UCA. We denote by $R_{tn}$, $\cdots$, and $R_{t1}$ the radii of $n$D transmit QF-UCA cells, $\cdots$, and 1D transmit UCA cell, respectively. We denote by $R_{rn}$, $\cdots$, and $R_{r1}$ the radii of $n$D receive QF-UCA cells, $\cdots$, and 1D receive UCA cell, respectively. Then, we denote by $R_{tE}$ and $R_{rE}$ the radii of the entire transmit and receive QF-UCA, respectively, where $R_{tE} = \sum_{n=1}^{N} R_{tn}$ and $R_{rE} = \sum_{n=1}^{N} R_{rn}$. Generally, we assume that $R_E = R_{tE} = R_{rE}$.

The $n$D QF-UCA cell or array-element indexed with $k_n$ ($k_n \in [0, K_n-1]$, $n \in [1,N]$) at the transmitter and those indexed with $v_n$ ($v_n \in [0, V_n-1]$, $n \in [1,N]$) at the receiver are uniformly distributed along the center of $n$D QF-UCA with $R_{tn}$ ($n\in[1,N]$) and $R_{rn}$ ($n\in[1,N]$), respectively. In addition, $R_{tn}$ and $R_{rn}$ are required to be the same. In the $n$D ($n \in [1, N]$) QF-UCA coordinate system, $x$-axis is set as the direction from the center of $n$D QF-UCA cell to the first $(n-1)$D QF-UCA cell ($k_n = 0$, $v_n = 0$) while $z$-axis is the centering normal line pointing to the directly opposite receive QF-UCA cell. Correspondingly, $y$-axis are decided by the right-hand spiral rule. Then, we denote by ($k_N$, $\cdots$, $k_1$) and ($v_N$, $\cdots$, $v_1$) the global index of the array-elements corresponding to the entire transmit and receive $N$D QF-UCA antenna, respectively. The parameters $\varphi_n = \frac{2\pi k_n}{K_n}(n \in [1, N])$ and $\theta_n = \frac{2\pi v_n}{V_n}(n \in [1, N])$ denote the azimuths within the $n$D QF-UCA cell at the transmitter and receiver, respectively.

### III. $N$-DIMENSIONAL QF-UCA BASED OAM MULTIPLEXING TRANSMISSION SCHEME

In order to implement orthogonal OAM-modes transmission based on $N$D QF-UCA, we investigate the OAM multiplexing transmission including $N$OM scheme and $N$OD scheme. Here, we focus on the scenario while the transmit and receive QF-UCA antennas are aligned with each other.

#### A. N-Dimensional IDFT Based OAM Modulation (NOM)

The modulated signal on the array-element indexed with $k_1$ of transmit 1D UCA, represented by $x_{k_1}$, can be derived as follows:

$$x_{k_1} = \sum_{l_1=0}^{K_1-1} \frac{s_{1\_l_1}}{\sqrt{K_1}} e^{j\frac{2\pi l_1 k_1}{K_1}}, \quad (1)$$

where $l_1$ ($l_1 \in [0, K_1-1]$) denotes the carried 1D OAM-modes, $K_1$ is the number of array-elements on the 1D UCA and the maximum number of OAM-modes corresponding to the 1D UCA, $k_1$ ($k_1 \in [0, K_1-1]$) is the index of array-element on the 1D UCA, and $s_{1\_l_1}$ is the signal corresponding to 1D UCA OAM-mode $l_1$. Then, with $\boldsymbol{s}_1 = \begin{bmatrix} s_{1\_0}, \cdots s_{1\_l_1}, \cdots s_{1\_(K_1-1)} \end{bmatrix}^T$, the transmit signal vector for 1D UCA, represented by $\boldsymbol{X}_1$, can be derived as follows:

$$\boldsymbol{X}_1 = [x_0, \cdots, x_{k_1}, \cdots, x_{K_1-1}]^T = \boldsymbol{W}_1 \boldsymbol{s}_1, \quad (2)$$

where $\boldsymbol{W}_1$ denotes a 1D IDFT modulation matrix of order $K_1$ and can be given as follows:

$$\boldsymbol{W}_1 = \begin{bmatrix} 1 & \cdots & 1 & \cdots & 1 \\ \vdots & \ddots & \vdots & \ddots & \vdots \\ 1 & \cdots & e^{j\frac{2\pi l_1 k_1}{K_1}} & \cdots & e^{j\frac{2\pi(K_1-1)k_1}{K_1}} \\ \vdots & \ddots & \vdots & \ddots & \vdots \\ 1 & \cdots & e^{j\frac{2\pi l_1(K_1-1)}{K_1}} & \cdots & e^{j\frac{2\pi(K_1-1)^2}{K_1}} \end{bmatrix}. \quad (3)$$

Then, for 2D QF-UCA, the modulated signal corresponding to the $k_1$th array-element of the $k_2$th ($k_2 \in [0, K_2-1]$) UCA cell, represented by $x_{(k_2, k_1)}$, can be given as follows:

$$x_{(k_2, k_1)} = \sum_{l_2=0}^{K_2-1} \sum_{l_1=0}^{K_1-1} \frac{s_{(l_2, l_1)}}{\sqrt{K_2 K_1}} e^{j\frac{2\pi l_1 k_1}{K_1}} e^{j\frac{2\pi l_2 k_2}{K_2}}, \quad (4)$$

where $s_{(l_2, l_1)}$ is the transmit signal corresponding to 1D UCA OAM-mode $l_1$ and 2D QF-UCA OAM-mode $l_2$, $k_2$ is the index of 1D UCA on the 2D QF-UCA, $K_2$ is the number of cells on the 2D QF-UCA, and $K_2 K_1$ is the maximum number of OAM-modes corresponding to the 2D QF-UCA. Then, with $\boldsymbol{s}_{2\_l_2} = \begin{bmatrix} s_{(l_2, 0)} \cdots s_{(l_2, l_1)} \cdots s_{(l_2, K_1-1)} \end{bmatrix}^T$, the transmit signal vector on the cell indexed with $k_2$ on the 2D QF-UCA, represented by $\boldsymbol{x}_{2\_k_2}$, can be derived as follows:

$$\boldsymbol{x}_{2\_k_2} = \sum_{l_2=0}^{K_2-1} \frac{e^{j\frac{2\pi l_2 k_2}{K_1}}}{\sqrt{K_2}} \boldsymbol{W}_1 \boldsymbol{s}_{2\_l_2}. \quad (5)$$

We define by $\boldsymbol{E}_1$ the identity matrix of order $K_1$, where the order is related to the number of 1D UCA array-elements $K_1$. Then, the 2D IDFT modulation matrix of order $K_2 K_1$, denoted by $\boldsymbol{W}_2$, can be given as follows:

$$\boldsymbol{W}_2 = \begin{bmatrix} \boldsymbol{E}_1 & \cdots & \boldsymbol{E}_1 & \cdots & \boldsymbol{E}_1 \\ \vdots & \ddots & \vdots & \ddots & \vdots \\ \boldsymbol{E}_1 & \cdots & e^{j\frac{2\pi l_2 k_2}{K_2}}\boldsymbol{E}_1 & \cdots & e^{j\frac{2\pi(K_2-1)k_2}{K_2}}\boldsymbol{E}_1 \\ \vdots & \ddots & \vdots & \ddots & \vdots \\ \boldsymbol{E}_1 & \cdots & e^{j\frac{2\pi l_2(K_2-1)}{K_2}}\boldsymbol{E}_1 & \cdots & e^{j\frac{2\pi(K_2-1)^2}{K_2}}\boldsymbol{E}_1 \end{bmatrix}. \quad (6)$$

Then, we denote by $\boldsymbol{W}_{k_2}$ the 2D nested IDFT modulation matrix of order $K_2 K_1$, which can be given as follows:

$$\boldsymbol{W}_{k_2} = \boldsymbol{W}_2 \boldsymbol{\Lambda}_2$$
$$= \begin{bmatrix} \boldsymbol{W}_1 & \cdots & \boldsymbol{W}_1 & \cdots & \boldsymbol{W}_1 \\ \vdots & \ddots & \vdots & \ddots & \vdots \\ \boldsymbol{W}_1 & \cdots & e^{j\frac{2\pi l_2 k_2}{K_2}}\boldsymbol{W}_1 & \cdots & e^{j\frac{2\pi(K_2-1)k_2}{K_2}}\boldsymbol{W}_1 \\ \vdots & \ddots & \vdots & \ddots & \vdots \\ \boldsymbol{W}_1 & \cdots & e^{j\frac{2\pi l_2(K_2-1)}{K_2}}\boldsymbol{W}_1 & \cdots & e^{j\frac{2\pi(K_2-1)^2}{K_2}}\boldsymbol{W}_1 \end{bmatrix}. \quad (7)$$

The elements of $\boldsymbol{W}_{k_2}$ are related to the 1D IDFT modulation matrix $\boldsymbol{W}_1$ and are distinguished from those of $\boldsymbol{W}_2$. The notation $\boldsymbol{\Lambda}_2$ is the scalar block matrix of order $K_2 K_1$ and can be denoted as follows:

$$\boldsymbol{\Lambda}_2 = blkdiag(\boldsymbol{W}_{k_1}, \cdots, \boldsymbol{W}_{k_1}) = blkdiag(\boldsymbol{W}_1, \cdots, \boldsymbol{W}_1), \quad (8)$$

where 1D nested IDFT modulation matrix $\boldsymbol{W}_{k_1} = \boldsymbol{W}_1$ only in 1D UCA.

Thus, the transmit signal vector on 2D QF-UCA, represented by $\boldsymbol{X}_2$, can be derived as follows:

$$\boldsymbol{X}_2 = \begin{bmatrix} \boldsymbol{x}_{2\_0} \\ \vdots \\ \boldsymbol{x}_{2\_k_2} \\ \vdots \\ \boldsymbol{x}_{2\_K_2-1} \end{bmatrix} = \boldsymbol{W}_2 \begin{bmatrix} \boldsymbol{W}_1 \boldsymbol{s}_{2\_0} \\ \vdots \\ \boldsymbol{W}_1 \boldsymbol{s}_{2\_l_2} \\ \vdots \\ \boldsymbol{W}_1 \boldsymbol{s}_{2\_K_2-1} \end{bmatrix} = \boldsymbol{W}_{k_2} \begin{bmatrix} \boldsymbol{s}_{2\_0} \\ \vdots \\ \boldsymbol{s}_{2\_l_2} \\ \vdots \\ \boldsymbol{s}_{2\_K_2-1} \end{bmatrix}. \quad (9)$$

Then, we expand to high-dimensional OAM modulation based on high-dimensional QF-UCA. The high-order nested IDFT modulation matrices and scalar matrices are in the same form as in Eqs. (7) and (8), respectively, only with different orders. We define by $\boldsymbol{E}_{n-1}$ ($n \in (1, N]$) the identity matrix of order $\prod_{i=1}^{n-1} K_i$, whose order is related to the number of each dimensional QF-UCA cells. Then, the $n$D IDFT modulation matrix of order $\prod_{i=1}^{n} K_i$, denoted by $\boldsymbol{W}_n$, can be given as follows:

$$\boldsymbol{W}_n = \begin{bmatrix} \boldsymbol{E}_{n-1} & \cdots & \boldsymbol{E}_{n-1} & \cdots & \boldsymbol{E}_{n-1} \\ \vdots & \ddots & \vdots & \ddots & \vdots \\ \boldsymbol{E}_{n-1} & \cdots & e^{j\frac{2\pi l_n k_n}{K_n}} \boldsymbol{E}_{n-1} & \cdots & e^{j\frac{2\pi (K_n-1) k_n}{K_n}} \boldsymbol{E}_{n-1} \\ \vdots & \ddots & \vdots & \ddots & \vdots \\ \boldsymbol{E}_{n-1} & \cdots & e^{j\frac{2\pi l_n (K_n-1)}{K_n}} \boldsymbol{E}_{n-1} & \cdots & e^{j\frac{2\pi (K_n-1)^2}{K_n}} \boldsymbol{E}_{n-1} \end{bmatrix}. \quad (10)$$

We denote by $\boldsymbol{W}_{k_n}$ the $n$D nested IDFT modulation matrix of order $\prod_{i=1}^{n} K_i$, which can be given as follows:

$$\boldsymbol{W}_{k_n} = \boldsymbol{W}_n \boldsymbol{\Lambda}_n$$
$$= \begin{bmatrix} \boldsymbol{W}_{k_{n-1}} & \cdots & \boldsymbol{W}_{k_{n-1}} & \cdots & \boldsymbol{W}_{k_{n-1}} \\ \vdots & \ddots & \vdots & \ddots & \vdots \\ \boldsymbol{W}_{k_{n-1}} & \cdots & e^{\frac{j2\pi l_n k_n}{K_n}} \boldsymbol{W}_{k_{n-1}} & \cdots & e^{\frac{j2\pi (K_n-1) k_n}{K_n}} \boldsymbol{W}_{k_{n-1}} \\ \vdots & \ddots & \vdots & \ddots & \vdots \\ \boldsymbol{W}_{k_{n-1}} & \cdots & e^{\frac{j2\pi l_n (K_n-1)}{K_n}} \boldsymbol{W}_{k_{n-1}} & \cdots & e^{\frac{j2\pi (K_n-1)^2}{K_n}} \boldsymbol{W}_{k_{n-1}} \end{bmatrix}. \quad (11)$$

The elements of $\boldsymbol{W}_{k_n}$ are related to the $(n-1)$D nested IDFT modulation matrix $\boldsymbol{W}_{k_{n-1}}$. The notation $\boldsymbol{\Lambda}_n = blkdiag(\boldsymbol{W}_{k_{n-1}}, \cdots, \boldsymbol{W}_{k_{n-1}})$ is the scalar block matrix of order $\prod_{i=1}^{n} K_i$.

Thus, with $\boldsymbol{s}_{n\_l_n} = [s_{n-1\_0} \cdots, s_{n-1\_l_{n-1}} \cdots, s_{n-1\_K_{n-1}-1}]^T$ ($n \in (1, N]$), the transmit signal vector on the cell indexed with $k_n$ on the $n$D QF-UCA, represented by $\boldsymbol{x}_{n\_k_n}$, can be derived as follows:

$$\boldsymbol{x}_{n\_k_n} = \sum_{l_n=0}^{K_n-1} \frac{e^{j\frac{2\pi l_n k_n}{K_n}}}{\sqrt{K_n}} \boldsymbol{W}_{k_{n-1}} \boldsymbol{s}_{n\_l_n}. \quad (12)$$

Then, distinguishing the nested IDFT modulation matrix from other dimensions, the $(N-1)$D nested IDFT modulation matrix can be given as follows:

$$\boldsymbol{W}_{k_{N-1}} = \boldsymbol{W}_{N-1} \boldsymbol{Q}_{N-1} \boldsymbol{\Lambda}_{N-1}, \quad (13)$$

where $\boldsymbol{W}_{N-1}$ is the $(N-1)$D IDFT modulation matrix of order $\prod_{n=1}^{N-1} K_n$, $\boldsymbol{Q}_{N-1}$ is precoding matrix, $\boldsymbol{\Lambda}_{N-1}$ is the scalar block matrix of order $\prod_{n=1}^{N-1} K_n$.

Then, we will give the relevant signals in the $N$D QF-UCA. The modulated signal on array-element indexed with $(k_N, \cdots, k_1)$ on the $N$D QF-UCA, represented by $x_{(k_N, \cdots, k_1)}$, can be given as follows:

$$x_{(k_N, \cdots, k_1)} = \sum_{l_N=0}^{K_N-1} \cdots \sum_{l_1=0}^{K_1-1} \frac{s_{(l_N, \cdots, l_1)}}{\sqrt{K_N \cdots K_1}} e^{j\frac{2\pi l_1 k_1}{K_1}} \cdots e^{j\frac{2\pi l_N k_N}{K_N}}, \quad (14)$$

where $s_{(l_N, \cdots, l_1)}$ is the transmit signal corresponding to $N$D QF-UCA OAM-mode $l_N$ ($l_N \in [0, K_N - 1]$), $\cdots$, $n$D QF-UCA OAM-mode $l_n$ ($l_n \in [0, K_n - 1]$), $\cdots$, and 1D UCA OAM-mode $l_1$.

Thus, with $\boldsymbol{s}_{N\_l_N} = [s_{N-1\_0} \cdots, s_{N-1\_l_{N-1}} \cdots, s_{N-1\_K_{N-1}-1}]^T$, the transmit signal vector on the cell indexed with $k_N$ on the $N$D QF-UCA, represented by $\boldsymbol{x}_{N\_k_N}$, can be derived as follows:

$$\boldsymbol{x}_{N\_k_N} = \sum_{l_N=0}^{K_N-1} \frac{e^{j\frac{2\pi l_N k_N}{K_N}}}{\sqrt{K_N}} \boldsymbol{W}_{k_{N-1}} \boldsymbol{s}_{N\_l_N}. \quad (15)$$

Then, the $N$D IDFT modulation matrix of order $\prod_{n=1}^{N} K_n$ denoted by $\boldsymbol{W}_N$, can be given as follows:

$$\boldsymbol{W}_N = \begin{bmatrix} \boldsymbol{E}_{N-1} & \cdots & \boldsymbol{E}_{N-1} & \cdots & \boldsymbol{E}_{N-1} \\ \vdots & \ddots & \vdots & \ddots & \vdots \\ \boldsymbol{E}_{N-1} & \cdots & e^{j\frac{2\pi l_N k_N}{K_N}} \boldsymbol{E}_{N-1} & \cdots & e^{j\frac{2\pi (K_N-1) k_N}{K_N}} \boldsymbol{E}_{N-1} \\ \vdots & \ddots & \vdots & \ddots & \vdots \\ \boldsymbol{E}_{N-1} & \cdots & e^{j\frac{2\pi l_N (K_N-1)}{K_N}} \boldsymbol{E}_{N-1} & \cdots & e^{j\frac{2\pi (K_N-1)^2}{K_N}} \boldsymbol{E}_{N-1} \end{bmatrix}. \quad (16)$$

Then, we denote by $\boldsymbol{W}_{k_N}$ the $N$D nested IDFT modulation matrix of order $\prod_{n=1}^{N} K_n$, which can be given as follows:

$$\boldsymbol{W}_{k_N} = \boldsymbol{W}_N \boldsymbol{\Lambda}_N$$
$$= \begin{bmatrix} \boldsymbol{W}_{k_{N-1}} & \cdots & \boldsymbol{W}_{k_{N-1}} & \cdots & \boldsymbol{W}_{k_{N-1}} \\ \vdots & \ddots & \vdots & \ddots & \vdots \\ \boldsymbol{W}_{k_{N-1}} & \cdots & e^{\frac{j2\pi l_N k_N}{K_N}} \boldsymbol{W}_{k_{N-1}} & \cdots & e^{\frac{j2\pi (K_N-1) k_N}{K_N}} \boldsymbol{W}_{k_{N-1}} \\ \vdots & \ddots & \vdots & \ddots & \vdots \\ \boldsymbol{W}_{k_{N-1}} & \cdots & e^{\frac{j2\pi l_N (K_N-1)}{K_N}} \boldsymbol{W}_{k_{N-1}} & \cdots & e^{\frac{j2\pi (K_N-1)^2}{K_N}} \boldsymbol{W}_{k_{N-1}} \end{bmatrix}. \quad (17)$$

The elements of $\boldsymbol{W}_{k_N}$ are related to the $(N-1)$D nested IDFT modulation matrix $\boldsymbol{W}_{k_{N-1}}$. The notation $\boldsymbol{\Lambda}_N$ is the scalar block matrix of order $\prod_{n=1}^{N} K_n$, which can be denoted as follows:

$$\boldsymbol{\Lambda}_N = blkdiag(\boldsymbol{W}_{k_{N-1}}, \cdots, \boldsymbol{W}_{k_{N-1}}). \quad (18)$$

Then, with $\boldsymbol{S}_N = [\boldsymbol{s}_{N\_0}, \cdots, \boldsymbol{s}_{N\_l_N}, \cdots, \boldsymbol{s}_{N\_K_N-1}]^T$, the transmit signal vector for $N$D QF-UCA, represented by $\boldsymbol{X}_N$, can be derived as follows:

$$\boldsymbol{X}_N = \boldsymbol{W}_N \begin{bmatrix} \boldsymbol{W}_{k_{N-1}} \boldsymbol{s}_{N\_0} \\ \vdots \\ \boldsymbol{W}_{k_{N-1}} \boldsymbol{s}_{N\_l_N} \\ \vdots \\ \boldsymbol{W}_{k_{N-1}} \boldsymbol{s}_{N\_K_N-1} \end{bmatrix} = \boldsymbol{W}_N \boldsymbol{\Lambda}_N \boldsymbol{S}_N. \quad (19)$$

### B. Wireless Channel Of N-Dimensional OAM Transmission

It is generally assumed that each dimensional QF-UCA cell corresponding to the transmitter and receiver satisfies $K_n = V_n$ ($n \in [1, N]$). In this paper, transmit signals are propagated in a LoS path.

According to the path loss of radio waves in free space, the expression of the complex channel gain between the array-element indexed with $(k_N, \cdots, k_1)$ on transmitter and the array-element indexed with $(v_N, \cdots, v_1)$ on receiver, denoted by $h_{(v_N,\cdots,v_1)}^{(k_N,\cdots,k_1)}$, can be given as follows:

$$h_{(v_N,\cdots,v_1)}^{(k_N,\cdots,k_1)} = \frac{\beta\lambda}{4\pi} \cdot \frac{e^{-j\frac{2\pi}{\lambda}d_{(v_N,\cdots,v_1)}^{(k_N,\cdots,k_1)}}}{d_{(v_N,\cdots,v_1)}^{(k_N,\cdots,k_1)}}, \quad (20)$$

where $\lambda$ represents the radio wavelength, $\beta$ denotes the parameter gathering relevant constants on antenna array-elements. The notation $d_{(v_N,\cdots,v_1)}^{(k_N,\cdots,k_1)}$ is the transmission distance from the array-element indexed with $(k_N, \cdots, k_1)$ on transmitter to the the array-element indexed with $(v_N, \cdots, v_1)$ on receiver, as derived in Eq. (21), where $D$ denotes the vertical distance between transmitter and receiver.

Then, we denote by $\boldsymbol{H}_{(v_N,\cdots,v_2)}^{(k_N,\cdots,k_2)}$ the 1D wireless channel gain matrix between the transmitter 1D UCA and receiver 1D UCA indexed $(k_N, \cdots, k_2)$ and $(v_N, \cdots, v_2)$, respectively. The elements of this matrix are the wireless channel gain between the array-elements on the 1D UCA at the corresponding indexes, which can be given as follows:

$$\boldsymbol{H}_{(v_N,\cdots,v_2)}^{(k_N,\cdots,k_2)} = \begin{bmatrix} h_{(v_N,\cdots,v_2,0)}^{(k_N,\cdots,k_2,0)} & \cdots & h_{(v_N,\cdots,v_2,0)}^{(k_N,\cdots,k_2,K_1-1)} \\ \vdots & \ddots & \vdots \\ h_{(v_N,\cdots,v_2,v_1)}^{(k_N,\cdots,k_2,0)} & \cdots & h_{(v_N,\cdots,v_2,v_1)}^{(k_N,\cdots,k_2,K_1-1)} \\ \vdots & \ddots & \vdots \\ h_{(v_N,\cdots,v_2,V_1-1)}^{(k_N,\cdots,k_2,0)} & \cdots & h_{(v_N,\cdots,v_2,V_1-1)}^{(k_N,\cdots,k_2,K_1-1)} \end{bmatrix}, \quad (22)$$

where the channel gain between the array-elements indexed $k_1$ and $v_1$ on the 1D UCA is $h_{(v_N,\cdots,v_1)}^{(k_N,\cdots,k_1)}$, has been given by Eq. (20).

Then, we denote by $\boldsymbol{H}_{(v_N,\cdots,v_{n+1},v_n)}^{(k_N,\cdots,k_{n+1},k_n)}$ the $(n{-}1)$D wireless channel gain matrix between the transmitter $(n{-}1)$D QF-UCA and receiver $(n{-}1)$D QF-UCA indexed $(k_N, \cdots, k_{n+1}, k_n)$ and $(v_N, \cdots, v_{n+1}, v_n)$, respectively. Thus, the $n$D wireless channel gain matrix between the transmitter $n$D UCA and receiver $n$D QF-UCA, denote by $\boldsymbol{H}_{(v_N,\cdots,v_{n+1})}^{(k_N,\cdots,k_{n+1})}$, can be given in Eq. (23).

Then, we denote by $\boldsymbol{H}_{(v_N)}^{(k_N)}$ the $(N{-}1)$D wireless channel gain matrix between the transmitter $(N{-}1)$D QF-UCA and receiver $(N{-}1)$D QF-UCA indexed $k_N$ and $v_N$, respectively. Thus, the $N$D wireless channel gain matrix between the transmitter $N$D QF-UCA and receiver $N$D QF-UCA, denote by $\boldsymbol{H}_N$, can be given as follows:

$$\boldsymbol{H}_N = \begin{bmatrix} \boldsymbol{H}_{(0)}^{(0)} & \cdots & \boldsymbol{H}_{(0)}^{(k_N)} & \cdots & \boldsymbol{H}_{(0)}^{(K_N-1)} \\ \vdots & \ddots & \vdots & \ddots & \vdots \\ \boldsymbol{H}_{(v_N)}^{(0)} & \cdots & \boldsymbol{H}_{(v_N)}^{(k_N)} & \cdots & \boldsymbol{H}_{(v_N)}^{(K_N-1)} \\ \vdots & \ddots & \vdots & \ddots & \vdots \\ \boldsymbol{H}_{(V_N-1)}^{(0)} & \cdots & \boldsymbol{H}_{(V_N-1)}^{(k_N)} & \cdots & \boldsymbol{H}_{(V_N-1)}^{(K_N-1)} \end{bmatrix}, \quad (24)$$

which is circulant block matrix, as indicated by the uniform circular cell structure of the QF-UCA at both the transmitter and receiver.

Based on Eqs. (20) and (14), we denote by $r_{(v_N,\cdots,v_1)}$ the received signal with the receiver array-element indexed $(v_N, \cdots, v_1)$, can be derived in Eq. (25), where we further denote the channel gain by $h_{(v_N,\cdots,v_1)}^{(l_N,\cdots,l_1)}$ as follows:

$$h_{(v_N,\cdots,v_1)}^{(l_N,\cdots,l_1)} = \sum_{k_N=0}^{K_N-1} \cdots \sum_{k_1=0}^{K_1-1} \frac{e^{j\frac{2\pi l_1 k_1}{K_1}} \cdots e^{j\frac{2\pi l_N k_N}{K_N}}}{\sqrt{K_N \cdots K_1}} h_{(v_N,\cdots,v_1)}^{(k_N,\cdots,k_1)}. \quad (26)$$

When the receiver array-element index and each dimensional OAM-modes are given, the amplitude and phase changes of the signal at the receiver can be represented.

$$\begin{aligned}
d_{(v_N,\cdots,v_1)}^{(k_N,\cdots,k_1)} &= \left\{ \begin{aligned} &D^2 + [(R_{rN}\cos\theta_N + \cdots + R_{r1}\cos\theta_1) - (R_{tN}\cos\varphi_N + \cdots + R_{t1}\cos\varphi_1)]^2 \\ &+ [(R_{rN}\sin\theta_N + \cdots + R_{r1}\sin\theta_1) - (R_{tN}\sin\varphi_N + \cdots + R_{t1}\sin\varphi_1)]^2 \end{aligned} \right\}^{\frac{1}{2}} \\
&= \left\{ D^2 + [R_N(\cos\theta_N - \cos\varphi_N) + \cdots + R_1(\cos\theta_1 - \cos\varphi_1)]^2 + [R_N(\sin\theta_N - \sin\varphi_N) + \cdots + R_1(\sin\theta_1 - \sin\varphi_1)]^2 \right\}^{\frac{1}{2}} \\
&= \left\{ \begin{aligned} &D^2 + R_N^2(\cos\theta_N - \cos\varphi_N)^2 + \cdots + R_1^2(\cos\theta_1 - \cos\varphi_1)^2 + [2R_1R_2(\cos\theta_1 - \cos\varphi_1)(\cos\theta_2 - \cos\varphi_2) + \cdots \\ &+ 2R_{N-1}R_N(\cos\theta_{N-1} - \cos\varphi_{N-1})(\cos\theta_N - \cos\varphi_N)] + [R_N^2(\sin\theta_N - \sin\varphi_N)^2 + \cdots + R_1^2(\sin\theta_1 - \sin\varphi_1)^2] \\ &+ [2R_1R_2(\sin\theta_1 - \sin\varphi_1)(\sin\theta_2 - \sin\varphi_2) + \cdots + 2R_{N-1}R_N(\sin\theta_{N-1} - \sin\varphi_{N-1})(\sin\theta_N - \sin\varphi_N)] \end{aligned} \right\}^{\frac{1}{2}} \\
&= \left\{ \begin{aligned} &D^2 + [2R_N^2(1-\cos(\theta_N-\varphi_N)) + \cdots + 2R_1^2(1-\cos(\theta_1-\varphi_1))] + [2R_1R_2(\cos(\theta_1-\theta_2) \\ &+ \cos(\varphi_1-\varphi_2) + \cos(\theta_1-\varphi_2) + \cos(\varphi_1-\theta_2)) + \cdots + 2R_{N-1}R_N(\cos(\theta_{N-1}-\theta_N) \\ &+ \cos(\varphi_{N-1}-\varphi_N) + \cos(\theta_{N-1}-\varphi_N) + \cos(\varphi_{N-1}-\theta_N))] \end{aligned} \right\}^{\frac{1}{2}} \\
&= \left\{ D^2 + \sum_{n=1}^{N} [2R_n^2(1-\cos(\theta_n-\varphi_n))] + \sum_{i=1}^{N-1}\sum_{j=i+1}^{N} \begin{aligned} &[2R_iR_j(\cos(\theta_i-\theta_j) + \cos(\varphi_i-\varphi_j) \\ &+ \cos(\theta_i-\varphi_j) + \cos(\varphi_i-\theta_j))]\end{aligned} \right\}^{\frac{1}{2}} \\
&= D\left\{ 1 + \frac{1}{D^2}\sum_{n=1}^{N} [2R_n^2(1-\cos(\theta_n-\varphi_n))] + \frac{1}{D^2}\sum_{i=1}^{N-1}\sum_{j=i+1}^{N} \begin{aligned}&[2R_iR_j(\cos(\theta_i-\theta_j) + \cos(\varphi_i-\varphi_j) \\ &+ \cos(\theta_i-\varphi_j) + \cos(\varphi_i-\theta_j))]\end{aligned} \right\}^{\frac{1}{2}} \\
&\approx D + \frac{1}{D}\sum_{n=1}^{N} [R_n^2(1-\cos(\theta_n-\varphi_n))] + \frac{1}{D}\sum_{i=1}^{N-1}\sum_{j=i+1}^{N} [R_iR_j(\cos(\theta_i-\theta_j) + \cos(\varphi_i-\varphi_j) + \cos(\theta_i-\varphi_j) + \cos(\varphi_i-\theta_j))].
\end{aligned} \quad (21)$$

$$\boldsymbol{H}^{(k_N,\cdots,k_{n+1})}_{(v_N,\cdots,v_{n+1})} = \begin{bmatrix} \boldsymbol{H}^{(k_N,\cdots,k_{n+1},0)}_{(v_N,\cdots,v_{n+1},0)} & \cdots & \boldsymbol{H}^{(k_N,\cdots,k_{n+1},k_n)}_{(v_N,\cdots,v_{n+1},0)} & \cdots & \boldsymbol{H}^{(k_N,\cdots,k_{n+1},K_n-1)}_{(v_N,\cdots,v_{n+1},0)} \\ \vdots & \ddots & \vdots & \ddots & \vdots \\ \boldsymbol{H}^{(k_N,\cdots,k_{n+1},0)}_{(v_N,\cdots,v_{n+1},v_n)} & \cdots & \boldsymbol{H}^{(k_N,\cdots,k_{n+1},k_n)}_{(v_N,\cdots,v_{n+1},v_n)} & \cdots & \boldsymbol{H}^{(k_N,\cdots,k_{n+1},K_n-1)}_{(v_N,\cdots,v_{n+1},v_n)} \\ \vdots & \ddots & \vdots & \ddots & \vdots \\ \boldsymbol{H}^{(k_N,\cdots,k_{n+1},0)}_{(v_N,\cdots,v_{n+1},V_n-1)} & \cdots & \boldsymbol{H}^{(k_N,\cdots,k_{n+1},k_n)}_{(v_N,\cdots,v_{n+1},V_n-1)} & \cdots & \boldsymbol{H}^{(k_N,\cdots,k_{n+1},K_n-1)}_{(v_N,\cdots,v_{n+1},V_n-1)} \end{bmatrix}, \quad (23)$$

$$\begin{aligned} r_{(v_N,\cdots,v_1)} &= \sum_{k_N=0}^{K_N-1}\cdots\sum_{k_1=0}^{K_1-1} x_{(k_N,\cdots,k_1)} h^{(k_N,\cdots,k_1)}_{(v_N,\cdots,v_1)} \\ &= \sum_{k_N=0}^{K_N-1}\cdots\sum_{k_1=0}^{K_1-1}\sum_{l_N=0}^{K_N-1}\cdots\sum_{l_1=0}^{K_1-1} \frac{s_{(l_N,\cdots,l_1)}}{\sqrt{K_N\cdots K_1}} e^{j\frac{2\pi l_1 k_1}{K_1}} \cdots e^{j\frac{2\pi l_N k_N}{K_N}} h^{(k_N,\cdots,k_1)}_{(v_N,\cdots,v_1)} \\ &= \sum_{l_N=0}^{K_N-1}\cdots\sum_{l_1=0}^{K_1-1} s_{(l_N,\cdots,l_1)} h^{(l_N,\cdots,l_1)}_{(v_N,\cdots,v_1)}, \end{aligned} \quad (25)$$

### C. N-Dimensional DFT Based OAM Demodulation(NOD)

We can derive the received signal vector $\boldsymbol{R}_N$ corresponding to $N$D QF-UCA cells through the LOS channels as follows:

$$\boldsymbol{R}_N = \boldsymbol{H}_N \boldsymbol{W}_N \boldsymbol{\Lambda}_N \boldsymbol{S}_N. \quad (27)$$

The signal vector after the $N$D QF-UCA based OAM demodulation, denoted by $\boldsymbol{\mathscr{R}}_N$, can be derived as follows:

$$\boldsymbol{\mathscr{R}}_N = \boldsymbol{W}_N^H \boldsymbol{R}_N = \boldsymbol{W}_N^H \boldsymbol{H}_N \boldsymbol{W}_N \boldsymbol{\Lambda}_N \boldsymbol{S}_N = \boldsymbol{\mathscr{H}}_N \boldsymbol{\Lambda}_N \boldsymbol{S}_N, \quad (28)$$

where $\boldsymbol{W}_N^H$ is the Hermitian matrix of $\boldsymbol{W}_N$, defined as the $N$D DFT demodulation matrix. For the circulant block matrix $\boldsymbol{H}_N$, there is a good property in connection with unitary decomposition that the unitary similarity transformation for $\boldsymbol{H}_N$ produces a diagonal block matrix. Specifically, using $N$D IDFT matrix $\boldsymbol{W}_N$ and $N$D DFT matrix $\boldsymbol{W}_N^H$, we have $\boldsymbol{W}_N^H \boldsymbol{H}_N \boldsymbol{W}_N = \boldsymbol{\mathscr{H}}_N$, where $\boldsymbol{\mathscr{H}}_N$ is a diagonal block matrix and can be derived as follows:

$$\boldsymbol{\mathscr{H}}_N = blkdiag(\sqrt{K_N}\left[\boldsymbol{H}^{(0)}_{(0)},\cdots,\boldsymbol{H}^{(K_N-1)}_{(0)}\right]\boldsymbol{W}_N). \quad (29)$$

We denote by $\boldsymbol{\mathscr{H}}_{N-1}$ the $k_N$th element on the diagonal of $\boldsymbol{\mathscr{H}}_N$. We name the above-mentioned operation the $N$D QF-UCA based OAM demodulation.

Then, performing a dimensionality reduction operation on Eq. (28). We take the $k_N$th row of $\boldsymbol{\mathscr{R}}_N$ and denote as $\boldsymbol{\mathscr{R}}_N(k_N)$. Further, we write $\boldsymbol{\mathscr{R}}_N(k_N)$ as $\tilde{\boldsymbol{R}}_{N-1}$ to denote the signal vector corresponding to the $(N–1)$D QF-UCA cells, which can be given as follows:

$$\begin{aligned} \tilde{\boldsymbol{R}}_{N-1} &= \boldsymbol{\mathscr{R}}_N(k_N) = \boldsymbol{\mathscr{H}}_{N-1} \boldsymbol{W}_{k_{N-1}} \tilde{\boldsymbol{S}}_{N-1} \\ &= \boldsymbol{\mathscr{H}}_{N-1} \boldsymbol{W}_{N-1} \boldsymbol{Q}_{N-1} \boldsymbol{\Lambda}_{N-1} \tilde{\boldsymbol{S}}_{N-1}. \end{aligned} \quad (30)$$

The signal vector after the $(N-1)$D QF-UCA based OAM demodulation, denoted by $\boldsymbol{\mathscr{R}}_{N-1}$, can be derived as follows:

$$\boldsymbol{\mathscr{R}}_{N-1} = \boldsymbol{U}_{N-1}^H \boldsymbol{P}_{N-1} \boldsymbol{W}_{N-1}^H \tilde{\boldsymbol{R}}_{N-1}, \quad (31)$$

where $\boldsymbol{W}_{N-1}^H$ is defined as the $(N-1)$D DFT demodulation matrix. $\boldsymbol{P}_{N-1}$ is the phase shift matrix, which converts the antisymmetric elements of the equivalent channel matrix into symmetric elements by easily inverting the signal phase to obtain the symmetric matrix $\boldsymbol{\mathscr{S}}_{N-1}$, which can be given as follows:

$$\begin{aligned} \boldsymbol{\mathscr{S}}_{N-1} &= \boldsymbol{P}_{N-1} \boldsymbol{W}_{N-1}^H \boldsymbol{\mathscr{H}}_{N-1} \boldsymbol{W}_{N-1} \\ &= \boldsymbol{U}_{N-1} \boldsymbol{V}_{N-1} \boldsymbol{Q}_{N-1}^H, \end{aligned} \quad (32)$$

where the symmetric matrix $\boldsymbol{\mathscr{S}}_{N-1}$ can be easily diagonalized, the Hermitian matrix of $\boldsymbol{U}_{N-1}$ is decoding matrix, $\boldsymbol{Q}_{N-1}^H$ is the Hermitian matrix of the precoding matrix $\boldsymbol{Q}_{N-1}$. We denote by $\boldsymbol{V}_N$ a diagonal block matrix with the element of the $k_N$th row as $\boldsymbol{V}_{N-1}$ and the remaining elements can be obtained from the $\boldsymbol{\mathscr{R}}_N(k_N)$ ($k_N \in [0, K_N-1]$) corresponding to Eq. (30). In addition, the lowest dimensional element of $\boldsymbol{V}_N$ corresponding to 1D transmit signal $s_{(l_N,\cdots,l_1)}$ can be denoted by $v_{(l_N,\cdots,l_1)}$, which denotes the singular values of equivalent channel matrix corresponding to $s_{(l_N,\cdots,l_1)}$.

Then, according to Eq. (32), we substitute Eq. (30) into Eq. (31) and futher rewrite $\boldsymbol{\mathscr{R}}_{N-1}$ as follows:

$$\begin{aligned} \boldsymbol{\mathscr{R}}_{N-1} &= \boldsymbol{U}_{N-1}^H \boldsymbol{U}_{N-1} \boldsymbol{V}_{N-1} \boldsymbol{Q}_{N-1}^H \boldsymbol{Q}_{N-1} \boldsymbol{\Lambda}_{N-1} \tilde{\boldsymbol{S}}_{N-1} \\ &= \boldsymbol{V}_{N-1} \boldsymbol{\Lambda}_{N-1} \tilde{\boldsymbol{S}}_{N-1}. \end{aligned} \quad (33)$$

We name the above-mentioned operation the $(N–1)$D QF-UCA based OAM demodulation. Then, a dimensionality reduction operation on Eq. (33) is performed. The $k_{N-1}$th row of $\boldsymbol{\mathscr{R}}_{N-1}$ is denoted as $\boldsymbol{\mathscr{R}}_{N-1}(k_{N-1})$. Further, we write $\boldsymbol{\mathscr{R}}_{N-1}(k_{N-1})$ as $\tilde{\boldsymbol{R}}_{N-2}$ to denote the signal vector corresponding to the $(N–2)$D QF-UCA cells, which can be given as follows:

$$\begin{aligned} \tilde{\boldsymbol{R}}_{N-2} &= \boldsymbol{\mathscr{R}}_{N-1}(k_{N-1}) \\ &= \boldsymbol{V}_{N-2} \boldsymbol{W}_{k_{N-2}} \tilde{\boldsymbol{S}}_{N-2} \\ &= \boldsymbol{V}_{N-2} \boldsymbol{W}_{N-2} \boldsymbol{\Lambda}_{N-2} \tilde{\boldsymbol{S}}_{N-2}. \end{aligned} \quad (34)$$

The signal vector after the $(N–2)$D QF-UCA based OAM

demodulation, denoted by $\mathscr{R}_{N-2}$, can be derived as follows:

$$\begin{aligned}\mathscr{R}_{N-2} &= \boldsymbol{W}_{N-2}^H \boldsymbol{V}_{N-2}^{-1} \tilde{\boldsymbol{R}}_{N-2} \\ &= \boldsymbol{W}_{N-2}^H \boldsymbol{V}_{N-2}^{-1} \boldsymbol{V}_{N-2} \boldsymbol{W}_{N-2} \boldsymbol{\Lambda}_{N-2} \tilde{\boldsymbol{S}}_{N-2} \\ &= \boldsymbol{\Lambda}_{N-2} \tilde{\boldsymbol{S}}_{N-2}.\end{aligned} \quad (35)$$

We name the above-mentioned operation the $(N-2)$D OAM demodulation based on $(N-2)$D QF-UCA.

Then, we progress to low-dimensional QF-UCA demodulation. Performing a dimensionality reduction operation and taking the $k_3$th row of $\mathscr{R}_3$, we derive the signal vector $\tilde{\boldsymbol{R}}_2$ corresponding to 2D QF-UCA cells, which can be given as follows:

$$\tilde{\boldsymbol{R}}_2 = \mathscr{R}_3(k_3) = \boldsymbol{W}_{k_2} \tilde{\boldsymbol{S}}_2 = \boldsymbol{W}_2 \boldsymbol{\Lambda}_2 \tilde{\boldsymbol{S}}_2. \quad (36)$$

The signal vector after the 2D QF-UCA based OAM demodulation, denoted by $\mathscr{R}_2$, can be derived as follows:

$$\mathscr{R}_2 = \boldsymbol{W}_2^H \tilde{\boldsymbol{R}}_2 = \boldsymbol{W}_2^H \boldsymbol{W}_2 \boldsymbol{\Lambda}_2 \tilde{\boldsymbol{S}}_2 = \boldsymbol{\Lambda}_2 \tilde{\boldsymbol{S}}_2. \quad (37)$$

Next, performing a dimensionality reduction operation and taking the $k_2$th row of $\mathscr{R}_2$, we can derive the signal vector $\tilde{\boldsymbol{R}}_1$ corresponding to 1D UCA cells, which can be given as follows:

$$\tilde{\boldsymbol{R}}_1 = \mathscr{R}_2(k_2) = \boldsymbol{W}_{k_1} \tilde{\boldsymbol{s}}_1 = \boldsymbol{W}_1 \tilde{\boldsymbol{s}}_1. \quad (38)$$

The signal vector after the 1D UCA based OAM demodulation, denoted by $\mathscr{R}_1$, can be derived as follows:

$$\mathscr{R}_1 = \boldsymbol{W}_1^H \tilde{\boldsymbol{R}}_1 = \boldsymbol{W}_1^H \boldsymbol{W}_1 \tilde{\boldsymbol{s}}_1 = \tilde{\boldsymbol{s}}_1. \quad (39)$$

### D. Exploring Optimal SE With Optimal Array-element Layout And Multiplexing Transmission Dimension

The SE of our proposed $N$D QF-UCA based OAM multiplexing transmission system can be derived as follows:

$$SE = \sum_{l_N=0}^{K_N-1} \cdots \sum_{l_1=0}^{K_1-1} \log_2\left(1 + \frac{\left|v_{(l_N,\cdots,l_1)}\right|^2 \left|s_{(l_N,\cdots,l_1)}\right|^2}{\sigma^2(l_N,\cdots,l_1)}\right), \quad (40)$$

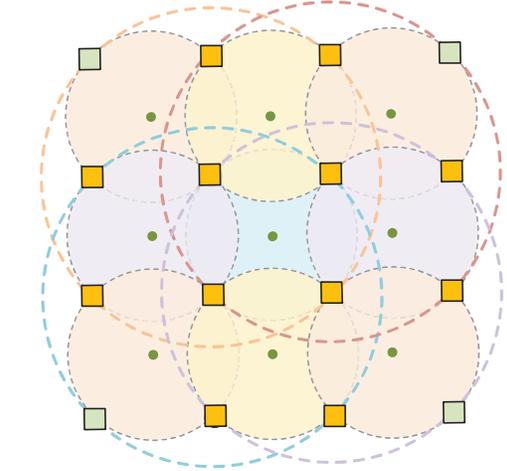

(a) Array-element layout type 1: each UCA cell in one dimension shares the same one cell in the center.

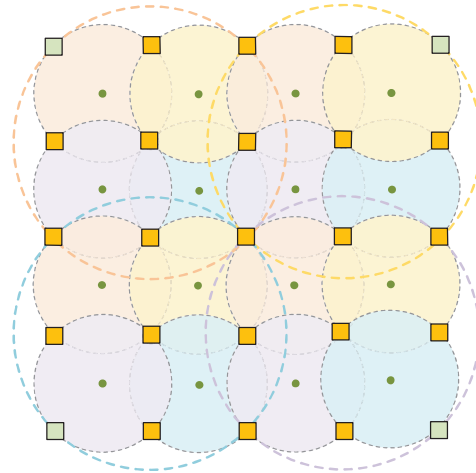

(b) Array-element layout type 2: $n$D adjacent UCA cells share the same cell in the center while $(n+ra)$D adjacent UCA cells are intersecting.

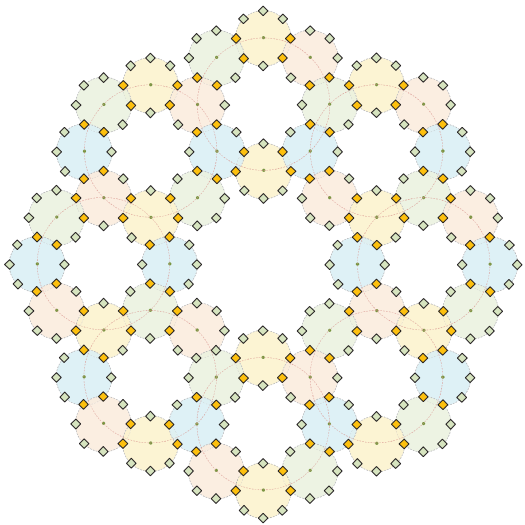

(c) Array-element layout type 3: adjacent UCA cells in each dimension are intersecting.

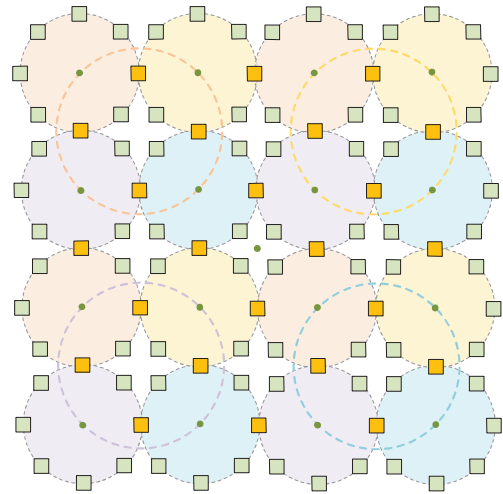

(d) Array-element layout type 4: adjacent UCA cells in each dimension are tangential.

Fig. 3. The different array-element layout types.

where $\sigma^2(l_N, \cdots, l_1)$ denote the noise variance corresponding to the OAM-modes indexed $(l_N, \cdots, l_1)$, $v_{(l_N,\cdots,l_1)}$ denote the singular values of equivalent channel matrix corresponding to $s_{(l_N,\cdots,l_1)}$.

For our proposed $N$D QF-UCA based OAM multiplexing transmission system, both the array-element layout and the modulation dimension affect the SE of the OAM multiplexing transmission system. The SE of our proposed system can be optimized by selecting the optimal array-element layout and multiplexing transmission dimension. Designing the layouts of array-elements and multiplexing transmission dimensions to facilitate the transmission of a greater number of orthogonal OAM-modes is an intriguing question that merits thorough investigation.

The various array-element layouts can be summarised in the types of array-element layouts shown in Fig. 3. Taking the 3D QF-UCA as an example, Fig. 3(a) is the array-element layout type where each UCA cell in one dimension shares the same cell in the center (referred to as array-element layout type 1). Fig. 3(b) is the array-element layout type where $n$D adjacent UCA cells share the same central cell while $(n+ra)$D adjacent UCA cells are intersecting (referred to as array-element layout type 2), where $ra \in [1, N-n]$ is a random positive integer. Fig. 3(c) is the array-element layout type where adjacent UCA cells in each dimension are intersecting (referred to as array-element layout type 3). Fig. 3(d) is the array-element layout type where adjacent UCA cells in each dimension are tangential (referred to as array-element layout type 4).

Considering the uniform distribution of array-elements in UCAs, specific criteria must be adhered to concerning the layout of QF-UCA antennas. Designing the $n$D QF-UCA, the radii $R_n$ and $R_{n-1}$ of the $n$th and $(n-1)$th dimensional UCA cells are required to satisfy certain predefined conditions. Futher, the numbers of $n$th and $(n-1)$th dimensional UCA cells $K_n$ and $K_{n-1}$ also are required to satisfy certain predefined conditions. The conditions corresponding to array-element layout types 1, 2, 3, and 4, can be given in Eqs. (41), (42), (43), and (44) respectively. The parameters $n+1 \in (1, N]$, $i_{2n} \in [1, K_n)$, $i_{2n-1} \in [1, K_n)$, and $i_n \in [1, K_n)$ are positive integers.

Consequently, the problem of achieving optimal SE through an optimal layout of array-elements and multiplexing transmission dimension, denoted by $\boldsymbol{P}_1$, can be formulated as follows:

$$\boldsymbol{P}_1 : \max_{n \in [1,N]} SE(N, K_n)$$
$$= \sum_{l_N=0}^{K_N-1} \cdots \sum_{l_1=0}^{K_1-1} \log_2\left(1 + \frac{\left|v_{(l_N,\cdots,l_1)}\right|^2 \left|s_{(l_N,\cdots,l_1)}\right|^2}{\sigma^2(l_N, \cdots, l_1)}\right). \quad (45)$$

We develop the optimal array-element layout and multiplexed transmission dimension algorithm for $\boldsymbol{P}_1$. We search for array-element layout types (*type b*, $b \in [1,4]$) that satisfy the corresponding conditions. The parameters $N_r$ is the number of all array-elements and we fix $N_r = N_a$ to find the maximum SE with the same number of array-elements. Then, we employ an exhaustive search for all combinations of $K_n$ and $N$ under the condition that *type b* is satisfied to calculate the SE. After comparison, the optimal SE and the corresponding array-element layout are obtained. Using

$$type\ 1: \begin{cases} \dfrac{R_n}{R_{n+1}} = 1; \\ \cos \dfrac{\pi i_{2n-1}}{K_n} = \sin \dfrac{\pi}{K_{n+1}}; \\ K_n\left(\dfrac{1}{2} - \dfrac{1}{K_{n+1}}\right) = i_{2n-1} + i_{2n}. \end{cases} \quad (41)$$

$$type\ 2: \begin{cases} \dfrac{R_n}{R_{n+1}} = \begin{cases} 1, ra \neq 1; \\ \sin \dfrac{\pi}{K_{n+1}}, ra = 1; \end{cases} \\ \cos \dfrac{\pi i_{2n-1}}{K_n} = \dfrac{R_{n+1}}{R_n} \sin \dfrac{\pi}{K_{n+1}}; \\ K_n\left(\dfrac{1}{2} - \dfrac{1}{K_{n+1}}\right) = i_{2n-1} + i_{2n}. \end{cases} \quad (42)$$

$$type\ 3: \begin{cases} \sin \dfrac{\pi}{K_{n+1}} < \dfrac{R_n}{R_{n+1}} < 1; \\ \cos \dfrac{\pi i_{2n-1}}{K_n} = \dfrac{R_{n+1}}{R_n} \sin \dfrac{\pi}{K_{n+1}}; \\ K_n\left(\dfrac{1}{2} - \dfrac{1}{K_{n+1}}\right) = i_{2n-1} + i_{2n}. \end{cases} \quad (43)$$

$$type\ 4: \begin{cases} \dfrac{R_n}{R_{n+1}} = \sin \dfrac{\pi}{K_{n+1}}; \\ K_{n+1} = \dfrac{2K_n}{K_n - 2i_n}. \end{cases} \quad (44)$$

**Algorithm 1**, we can find the optimal array-element layout and multiplexing transmission dimension, which results in the maximum SE when the number of array-elements is fixed and the power is allocated on average to each OAM-mode.

In addition, when the power is uniformly distributed across each OAM-mode, we define the average SE of each array-element as the efficiency of array-element layouts (EOAL) in QF-UCA and utilize it to evaluate the array-element layouts in Section IV.

## IV. PERFORMANCE EVALUATIONS

In this section, numerical simulation results are presented to evaluate the performance of wireless communication for our developed $N$D QF-UCA based OAM multiplexing transmission scheme. Firstly, we explore different array-element layout types and their efficiency of array-element layout. Secondly, we compare SEs of the 4D, 3D, 2D, and 1D OAM multiplexing transmission as well as corresponding array-element layouts with the same array-elements number. Then, we compare SEs of the 4D, 3D, 2D, and 1D OAM multiplexing transmission for different transmission distances. Finally, we investigate different dimensional multiplexing transmission schemes based on the corresponding QF-UCA antenna with various array-element layouts. Throughout the evaluation, we set the communication system operating at the 5.8GHz frequency band and assume transceivers are aligned with each other. The constant $\beta$ is set to 1. The distance between the transmitter and receiver is 100m.

We explore the relationship among different array-element layout types and the efficiency of array-element layout in the same dimension OAM multiplexing transmission. In Fig. 4,

**Algorithm 1** : The optimal array-element layout and multiplexed transmission dimension for $\boldsymbol{P}_1$

1: Initialize $N_a$, $K_n = 1$, $n = 1$, $b = 1$, $\max SE = 0$, and $\frac{R_n}{R_{n+1}} \in (0, 1]$ where $n \in [1, N]$, $i_{2n} \in [1, K_n)$, $i_{2n-1} \in [1, K_n)$, and $i_n \in [1, K_n]$.
2: $K_n = K_n + 1$, $K_n \in [1, N_r]$, $n = n + 1$, $n \in [1, N]$.
3: Find the $K_n$ and $N$ that satisfies array-element layout conditions.
4: **if** satisfying $type\, b$ **and** $N_r = N_a$ **then**
5:     Calculating $SE$ by Eq. (45).
6:     **if** $SE < \max SE$ **then**
7:         Go to **Step 2**;
8:     **else**
9:         $\max SE = SE$;
10:        Go to **Step 2**;
11:     **end if**
12: **else if** $K_n < N_r$ **then**
13:     Go to **Step 2**;
14: **else**
15:     Go to **Step 17**;
16: **end if**
17: **while** $b \leq 4$ **do**
18:     $b = b + 1, K_n = 1, n = 1$; Go to **Step 4**;
19: **end while**
20: Return $\max SE$ and the corresponding $N$, $K_n$, and $R_n$.

the power is uniformly distributed across each OAM-mode, where antenna radius $R_E = 4$m. As shown in Fig. 4, the array-element layout type 1 exhibits the highest efficiency of array-element layout as compared with other layout types. The array-element layout type 2 is distinguished from other single types of layouts, which are a combination of two types of layouts and is not as efficient as type 1. The array-element layout type 3 and type 4 have lower efficiency of array-element layout. This is because the geometrical position between the circles in a fixed-size antenna relates to the number of UCAs that can be formed, which affects the number of transmit orthogonal data streams.

Fig. 5 depicts the SEs of the 4D, 3D, 2D, and 1D OAM multiplexing transmission and corresponding array-element layouts with the same array-elements number of 25. We design array-element layouts like array-element layout type 1 and type 2. The antenna radius $R_E = 4$m and the power is uniformly distributed across each OAM-mode. Compared with the traditional single-loop UCA with 25 array-elements (referred to as 1D UCA), the high-dimensional QF-UCA based OAM multiplexing transmission achieves higher SEs. We can obtain that the SE increases as the dimension of vortex multiplexing transmission increases without additional power and array-elements. This is because OAM multiplexing transmission based on high-dimensional QF-UCA antenna can provide more orthogonal OAM-modes than the number of array-elements. Also, the OAM beam of high-order modes diverge severely and OAM multiplexing transmission based on high-dimensional QF-UCA antenna can utilize a greater number of low-order OAM-modes. The results show that in addition to the traditional multiplexing of multiple antenna communication systems, exploring additional orthogonal multiplexing strategies holds promise. In addition, a comparison between Fig. 5(b) and Fig. 5(c) reveals an intriguing observation: the array-element layouts appear identical after design. However, Fig. 5(c) is a 4D QF-UCA with increased the shared of array-elements, facilitating the transmission of more orthogonal OAM-modes using the 4D OAM multiplexing transmission scheme. Designing the layouts of array-elements to facilitate the transmission of a greater number of orthogonal OAM-modes is an intriguing question that merits thorough investigation.

Figure 6 shows the SEs of the 4D, 3D, 2D, and 1D OAM multiplexing transmission considering different transmission distances. The array-element layouts are consistent with the corresponding layouts in Fig. 5. The SNR = 15dB and the power is allocated on average to each OAM-mode. As shown in Fig. 6, the traditional single-loop UCA with 25 array-elements (referred to as 1D UCA) exhibits lower SE as compared with other high-dimensional multiplexing transmission, and SE diminishes considerably over longer distances. This is because the high-order OAM-modes of the 1D OAM multiplexing transmission diverge severely with distance increasing, substantially decreasing the SE. However, high-dimensional OAM multiplexing transmission based on the 4D QF-UCA can transmit more low-order OAM-modes and still achieve relatively high SE at longer distances. We obtain that the high-dimensional OAM multiplexing transmission based on high-dimensional QF-UCA can significantly increase the transmission distance of the communication system without additional power and array-elements.

Figure 7 shows the EOAL of different dimensional multiplex transmission schemes based on the corresponding QF-UCA with various array-element layouts and different numbers of array-elements. It can be observed that for QF-UCA with different numbers of array-elements, the higher the dimensionality of QF-UCA achievable through array-element layout, the greater the efficiency of array-element layout in QF-UCA. This relationship stems from the increased multiplexing capability of OAM-modes achievable through high-dimensional QF-UCA in OAM multiplexing transmission. It also can be observed that for QF-UCA with the same number of array-elements (as indicated by legends such as 2D QF-UCA 25-element, 3D QF-UCA 25-element, and 4D QF-UCA 25-element), high-dimensional QF-UCA formed exhibit higher efficiency of array-element layout, consistent as shown in Fig. 5. For QF-UCA in the same dimension, such as the 2D QF-UCA with 9 array-element layout (denoted by legend 2D QF-UCA 9-element), it demonstrates superior efficiency of array-element layout as compared with the layout with 25 array-elements (denoted by legend 2D QF-UCA 25-element). This is due to the more effective utilization of shared array-elements, enabling the transmission of a greater number of OAM-modes with fewer array-elements. In addition, the high-dimensional QF-UCA is different from the traditional single-loop UCA, and the diverse array-element layout makes it possible for square arrays to generate vortex electromagnetic waves, which break through the traditional concept of generating vortex electromagnetic waves with uniform circular arrays. The above observations can be used to provide guidance for the design of the antenna size and the number of array-elements. When antenna size and the number of array-elements are determined, choosing the appropriate array-element layout and multiplexing transmission dimension achieves the maximum SE.

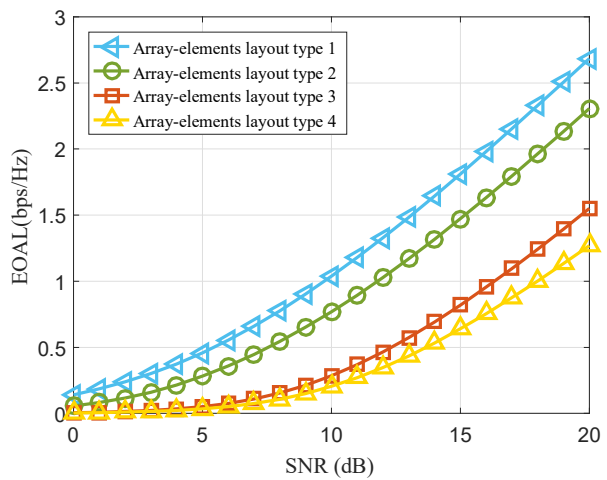

Fig. 4. Efficiency of array-element layout of different array-element layout types (array-element layout types as shown in Fig. 3(a), 3(b), 3(c), 3(d)).

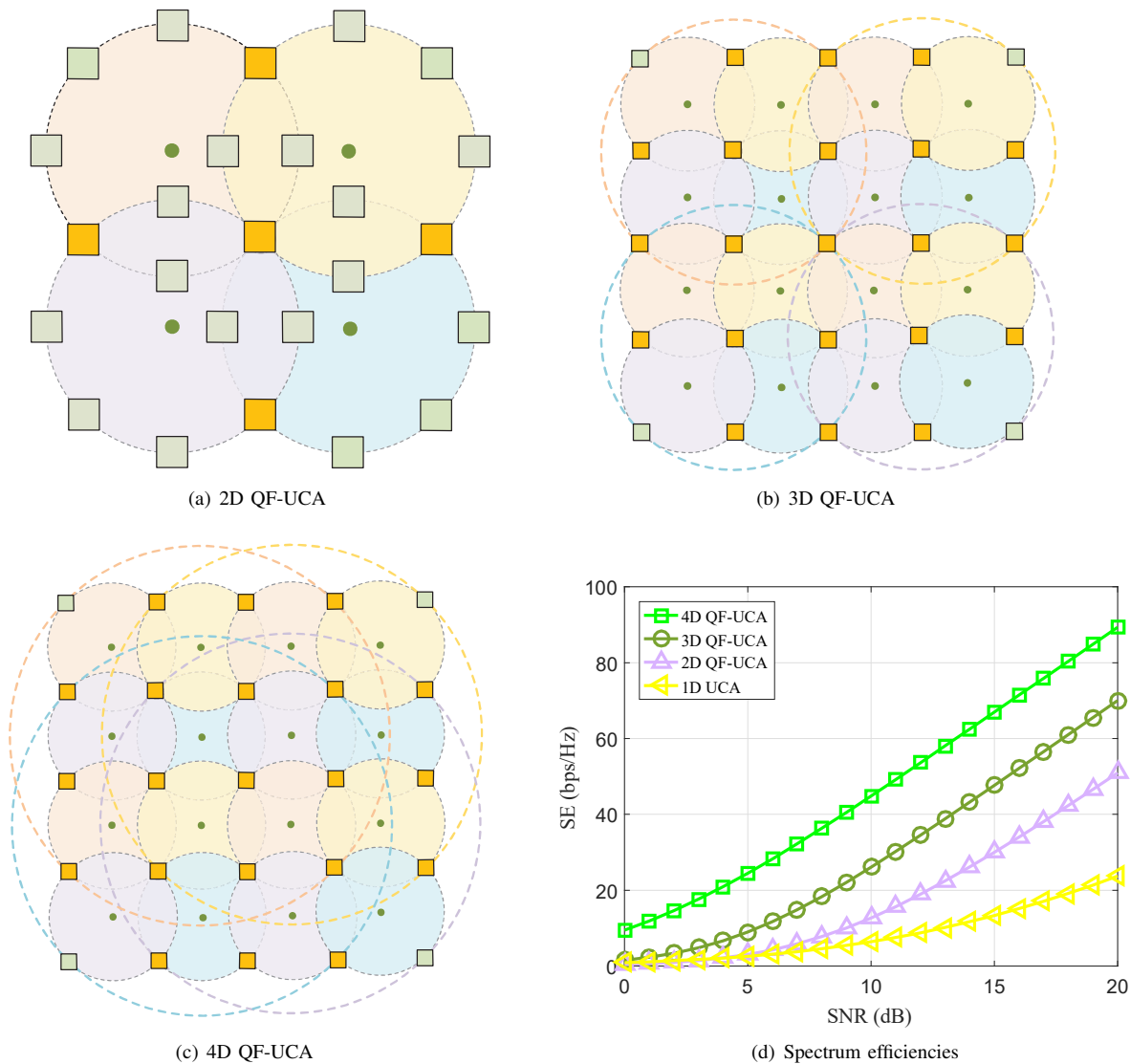

Fig. 5. Spectrum efficiencies of the 4D, 3D, 2D, and 1D OAM multiplexing transmission as well as corresponding array-element layouts with the same array-elements number of 25 (Fig. 5(a), Fig. 5(b), and Fig. 5(c) are the array-element layouts of 2D, 3D, and 4D QF-UCA antennas consisting of 25 array-elements, respectively).

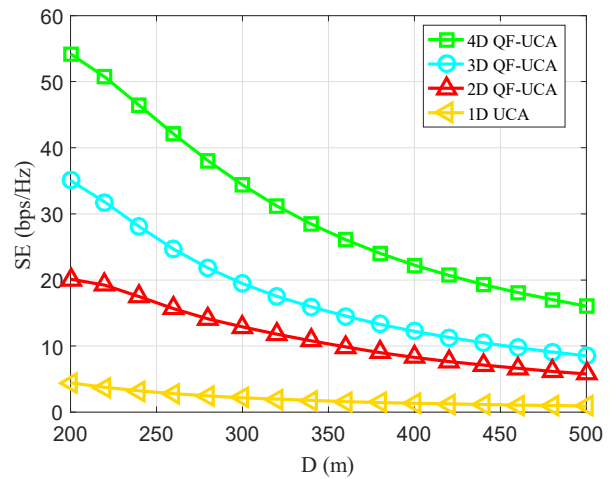

Fig. 6. Spectrum efficiencies of the 4D, 3D, 2D, and 1D OAM multiplexing transmission considering different transmission distances.

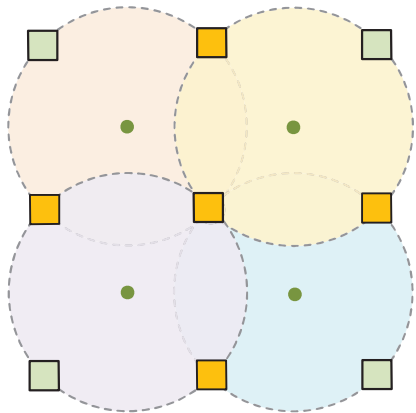

(a) 2D QF-UCA 9-element

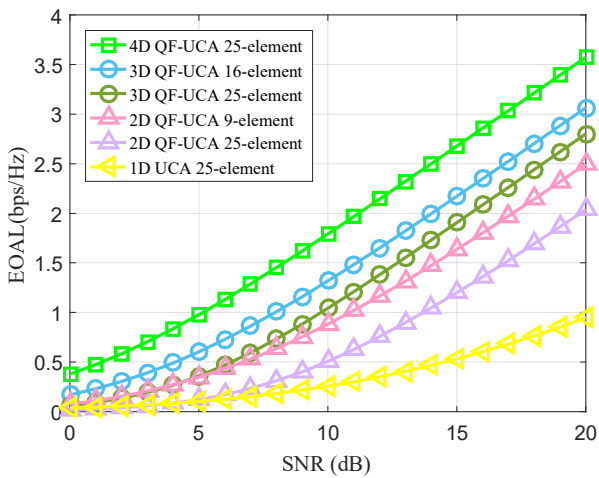

(b) Efficiency of array-element layout

Fig. 7. Efficiency of array-element layout of different dimensional multiplexing transmission schemes based on the corresponding QF-UCA with various array-element layouts and different numbers of array-elements (Fig. 3(a), Fig. 5(a), Fig. 5(b), Fig. 5(c), Fig. 7(a) are the array-element layouts of 3D QF-UCA 16-element, 2D QF-UCA 25-element, 3D QF-UCA 25-element, 4D QF-UCA 25-element, 2D QF-UCA 9-element, respectively).

## V. CONCLUSIONS

In this paper, we proposed the $N$D QF-UCA based OAM multiplexing transmission scheme, which combines the design of array-element layout and multiplexing transmission scheme, bringing a novel research direction to enhance the channel capacity. We designed $N$-dimensional QF-UCA antenna structure and compared different types of array-element layouts. Then, we developed the $N$-dimensional DFT based OAM modulation and demodulation schemes to transmit multiple OAM signals. Simulation results demonstrate that the optimal array-element layout of QF-UCA can be obtained using our proposed schemes to achieve maximum SE for OAM multiplexing transmission. The number of orthogonal data streams for the $N$-dimensional multiplexing transmission far exceeds the number of array-elements, providing a completely new solution to increase the number of available orthogonal streams when the antenna aperture and the number of array-elements are limited, which goes beyond the traditional concept of multiple antennas based wireless communications.